\documentclass{article}

\usepackage{amssymb}
\usepackage{amsmath}
\usepackage{tensor} % provides tensor index notation
\usepackage{commath} % Provides \dif, \norm, \abs, \del, \cbr, \sbr etc.
\usepackage{bm} % italic bold and greek bold
\usepackage{nomencl} % nomenclature
\makenomenclature

\usepackage[square,sort&compress]{natbib} % references

\usepackage{tikz}

\usepackage[normalem]{ulem} % allow strikethroughs (command \sout{})

\usepackage{authblk} % author affiliations package

\newcommand{\vol}{\operatorname{\mathfrak{v}}}
\newcommand{\VOL}{\operatorname{\mathfrak{V}}}
\DeclareMathOperator{\tr}{tr} % trace

\newcommand{\new}[1]{#1}
\newcommand{\old}[1]{}

\title{The Elastic Theory of Shells using Geometric Algebra}
\date{15 February 2017}
\author[1,*]{Alastair L Gregory}
\author[1]{Joan Lasenby}
\author[1]{Anurag Agarwal}
\affil[1]{Department of Engineering,
University of Cambridge,
Trumpington Street,
Cambridge, CB2 1PZ}
\affil[*]{\texttt{alg57@cam.ac.uk}}

\begin{document}

\maketitle

\begin{abstract}
  We present a novel derivation of the elastic theory of shells. We use the
  language of Geometric algebra, which allows us to express the fundamental laws
  in component-free form, thus aiding physical interpretation. It also provides
  the tools to express equations in an arbitrary coordinate system, which
  enhances their usefulness. The role of moments and angular velocity, and the
  apparent use by previous authors of an unphysical angular velocity, has been
  clarified through the use of a bivector representation. In the linearised
  theory, clarification of previous coordinate conventions which have been the
  cause of confusion, is provided, and the introduction of prior strain into the
  linearised theory of shells is made possible.
\end{abstract}

\section{Introduction}
Thin shells have been an active subject of research for some considerable time,
however, in attempting to understand the self excited oscillations of flexible
tubes, \old{the authors} \new{we} have had difficulties finding a complete and
rational theory in which the underlying physical principles are clear, and which
is easy to apply to the practical problem at hand. \new{Specifically, it was
  found that in order to have a full understanding of the assumptions of various
  shell theories, it was necessary to derive our own from first principles. We
  found that in doing this we were able to produce a theory with improved
  clarity, brevity, and with explicit results for linearisation about a deformed
  state.}

\old{The authors} \new{We} also have an interest in applying Geometric Algebra
(GA) \citep{Hestenes:1984vg,Lasenby:2000dh,Doran:2003jd} to new areas of the
physical sciences. GA provides the tools to formulate physical laws with as
little reference to coordinate systems as possible, which helps with the first
aim of clarifying the physical meaning of the equations produced, but it also
provides simple tools to allow these equations to be represented in arbitrary
coordinate systems, which ensures practical utility. This article aims to
provide, for the first time in this area, an introduction to shell theory using
GA. \new{While in this article we restrict the introduction of GA to the use of
  bivectors to represent torques and angular velocities, we hope that this will
  pave the way for more radical developments, such as those completed for the
  theory of rods \citep{McRobie:1999wv}.}

There are a large number (at least $10$) of linearised shell theories
\citep{Leissa:1973tw}. The derivations of these theories use a wide variety of
notations, coordinate systems, and conventions, making it very difficult to
compare the assumptions made. In addition, none of the theories reviewed by
\citep{Leissa:1973tw} allow for prior strain of the shell, which we wish to
include for our own analysis. More general shell theories have also been
produced, with the most extensive probably that by \citep{Naghdi:1972ul},
\new{which provides the basis for more modern works such as
  \citep{CiarletJr:2005vn,Antman:1995wm,Lacarbonara:2012bv},} though the theory
of \citep{Koiter:1966uk} has also been popular with some authors. While
rigorous, these theories have limited practical use. They generally require the
use of differential geometry \citep{CiarletJr:2005vn,Marsden:1994ty} whose
indicial expressions often hide much of the physical meaning of the equations.
\new{The general theory presented by \citep{Antman:1995wm} is relegated to a
  final chapter that does not stand alone, meaning that the entire book must be
  read to use the shell theory. \citep{Naghdi:1972ul} discusses in detail the
  different advantages of developing a shell theory directly from 3-dimensional
  elasticity or by considering 2-dimensional surfaces from the start.
  \citep{Antman:1995wm} restricts his development to the former, but we feel
  that a more concise and lucid theory can be obtained from the latter.}

We aim to \old{rectify this by using} \new{use} GA to develop an \old{clear}
accessible, \new{concise,} rational shell theory that can be easily linearised
to include pre-strain. In doing this we will provide new developments in the
representation of moments and angular velocities with bivectors, and in the
representation of bending, which is where most disagreements occur in linearised
shell theories.

\nomenclature[a-B]{$B$}{reference configuration}
\nomenclature[a-S]{$S$}{spatial configuration}
\nomenclature[g-phit]{$\phi_t$}{a motion of the reference configuration}
\nomenclature[a-t]{$t$}{time}
\nomenclature[a-X]{$X$}{a point in the reference configuration}
\nomenclature[a-Xi]{$\{X^i\}$}{coordinate system over the reference
  configuration}
\nomenclature[a-xi]{$\{x^i\}$}{convected coordinate system over the spatial
  configuration}
\nomenclature[a-Ei]{$\{E_i\}$}{frame for the tangent space of the reference
  configuration}
\nomenclature[a-ei]{$\{e_i\}$}{frame for the tangent space of the spatial
  configuration}
\nomenclature[a-Eir]{$\{E^i\}$}{reciprocal frame for the tangent space of the
  reference configuration}
\nomenclature[a-eir]{$\{e^i\}$}{reciprocal frame for the tangent space of the
  spatial configuration}
\nomenclature[a-I]{$I$}{local pseudoscalar on the reference configuration}
\nomenclature[a-i]{$i$}{local pseudoscalar on the spatial configuration}
\nomenclature[a-I3]{$I_3$}{pseudoscalar of 3-dimensional Euclidean space}
\nomenclature[a-E3]{$E_3$}{normal vector to the reference configuration}
\nomenclature[a-e3]{$e_3$}{normal vector to the spatial configuration}
\nomenclature[a-vol]{$\vol$}{volume form on the reference configuration}
\nomenclature[a-Vol]{$\VOL$}{volume form on the spatial configuration}
\nomenclature[x-nabla]{$\nabla$}{vector derivative}
\nomenclature[x-partial]{$\partial$}{vector derivative intrinsic to a surface}
\nomenclature[a-G]{$\mathsf G$}{metric, or first fundamental form, on the
  reference configuration}
\nomenclature[a-g]{$\mathsf g$}{metric, or first fundamental form, on the
  spatial configuration}
\nomenclature[a-B]{$\mathsf B$}{second fundamental form on the reference
  configuration}
\nomenclature[a-b]{$\mathsf b$}{second fundamental form on the spatial
  configuration}
\nomenclature[a-Ci]{$C_i$}{principal curvatures of the reference configuration}
\nomenclature[a-ci]{$c_i$}{principal curvatures of the spatial configuration}
\nomenclature[g-Gammaa]{$\Gamma^a_{ib}$}{Christoffel coefficients on the
  reference configuration}
\nomenclature[g-gammaa]{$\gamma^a_{ib}$}{Christoffel coefficients on the spatial
  configuration}
\nomenclature[a-EA]{$\{E_A\}$}{frame for bivectors on the reference
  configuration}
\nomenclature[a-eA]{$\{e_A\}$}{frame for bivectors on the spatial
  configuration}
\nomenclature[a-EAr]{$\{E^A\}$}{reciprocal frame for bivectors on the reference
  configuration}
\nomenclature[a-eAr]{$\{e^A\}$}{reciprocal frame for bivectors on the spatial
  configuration}
\nomenclature[g-GammaA]{$\Gamma^A_{iB}$}{bivector Christoffel coefficients on
  the reference configuration}
\nomenclature[g-gammaA]{$\gamma^A_{iB}$}{bivector Christoffel coefficients on
  the spatial configuration}
\nomenclature[g-lambdai]{$\lambda_i$}{principal stretches}
\nomenclature[a-F]{$\mathsf F$}{deformation gradient}
\nomenclature[a-C]{$\mathsf C$}{Cauchy-Green tensor}
\nomenclature[a-E]{$\mathsf E$}{Green-Lagrange strain tensor}
\nomenclature[a-H]{$\mathsf H$}{change of curvature tensor}
\nomenclature[a-V]{$V$}{velocity referred to the reference configuration}
\nomenclature[a-v]{$v$}{velocity referred to the spatial configuration}
\nomenclature[a-l]{$\mathsf l$}{strain rate tensor}
\nomenclature[a-n]{$\mathsf n$}{symmetric strain rate tensor}
\nomenclature[a-w]{$\mathsf w$}{antisymmetric strain rate tensor}
\nomenclature[a-Edot]{$\dot{\mathsf E}$}{rate of change of strain tensor}
\nomenclature[a-Hdot]{$\dot{\mathsf H}$}{rate of change of the change of
  curvature tensor}
\nomenclature[g-omega]{$\omega$}{angular velocity}
\nomenclature[a-b]{$b$}{body force per unit mass}
\nomenclature[a-c]{$c$}{body moments per unit mass}
\nomenclature[g-sigma]{$\sigma$}{Cauchy stress tensor}
\nomenclature[a-m]{$\mathsf m$}{couple-stress tensor}
\nomenclature[a-T]{$\mathsf T$}{first Piola-Kirchhoff stress tensor}
\nomenclature[a-M]{$\mathsf M$}{first reference couple-stress tensor}
\nomenclature[a-S]{$\mathsf S$}{second Piola-Kirchhoff stress tensor}
\nomenclature[a-N]{$\mathsf N$}{second reference couple-stress tensor}
\nomenclature[a-Mb]{$\bm{\mathsf M}$}{modified first reference couple-stress
  tensor}
\nomenclature[a-Nb]{$\bm{\mathsf N}$}{modified second reference couple-stress
  tensor}
\nomenclature[g-rho]{$\rho$}{area density of shell}
\nomenclature[g-rho0]{$\rho_0$}{time independent area density of shell}
\nomenclature[a-Stilde]{$\tilde{\mathsf S}$}{modified second Piola-Kirchhoff
  stress tensor}
\nomenclature[a-energy]{$e$}{internal energy per unit mass of the shell, defined
  on the spatial configuration}
\nomenclature[a-Energy]{$E$}{internal energy per unit mass of the shell, defined
  on the reference configuration}

\printnomenclature

\section{Geometry of Surfaces}
\begin{figure}
  \centering
  \includegraphics[width=10cm]{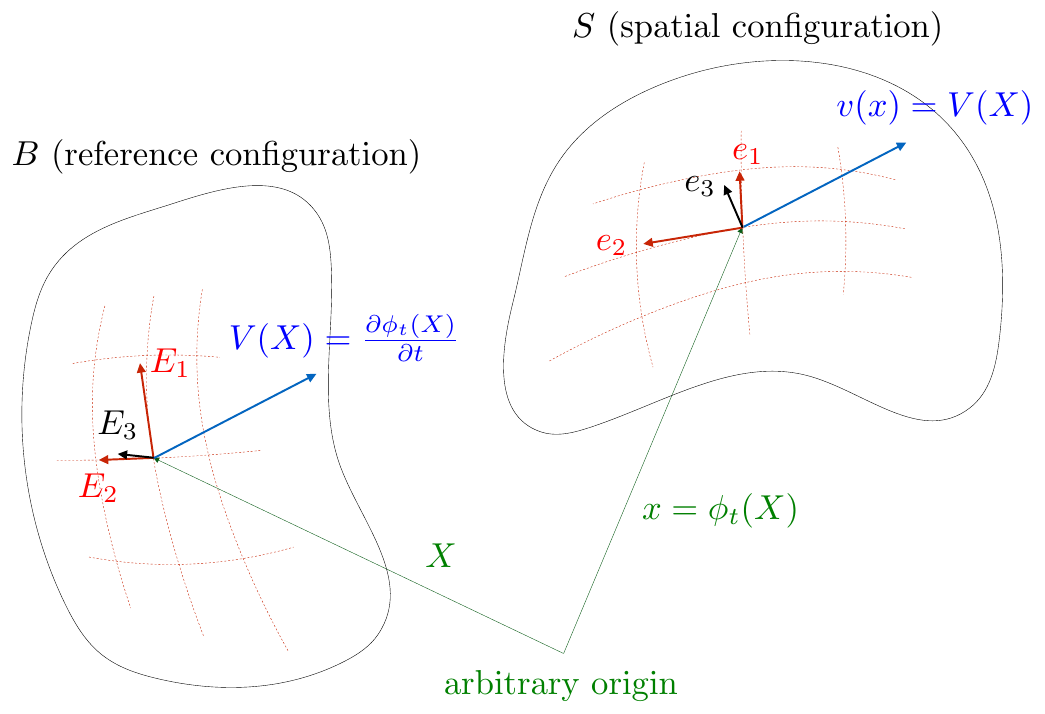}
  \caption{Surface geometry.}
  \label{fig:surfaceGeometry}
\end{figure}

Let $B$ and $S$ be $2$-dimensional surfaces embedded in $3$-dimensional
Euclidean space $\mathbb E^3$. $B$ is the {\bf reference configuration} of the
surface, and $S$ is the {\bf spatial configuration}, and the two are related by
the {\bf motion} $\phi_t$. At time $t$ the point $X\in B$ is moved to
$\phi_t(X)\in S$.  Let $\{X^i\}$ be coordinates over $B$, and $\{x^i\}$ be
coordinates over $S$. We follow the convention that the indices $i,j,k,\hdots$
run over $1,2$, and the indices $a,b,c,\hdots$ run over $1,2,3$. We restrict
$\{x^i\}$ to be {\bf convected coordinates} such that
$x^i(x)=X^i(\phi_t^{-1}(x))$ where $x\in S$. We denote the {\bf frame}
associated with $\{X^i\}$ by $E_i=\frac{\partial X}{\partial X^i}$, and
similarly, $e_i=\frac{\partial x}{\partial x^i}$. The {\bf reciprocal frames}
are denoted by $\{E^i\}$ and $\{e^i\}$, and are defined to satisfy $E^i\cdot
E_j=e^i\cdot e_j=\delta^i_j$. The frames on each configuration are illustrated
in \figref{fig:surfaceGeometry}.

The local {\bf pseudoscalars} in the reference and spatial configurations are
$I=\frac{E_1\wedge E_2}{\abs{E_1\wedge E_2}}$ and $i=\frac{e_1\wedge
e_2}{\abs{e_1\wedge e_2}}$, which satisfy $I^2=i^2=-1$. We denote the
pseudoscalar of $\mathbb E^3$ by $I_3$. We have defined orientations of both
configurations and $\mathbb E^3$ with these pseudoscalars, which allows us to
define the {\bf normal vectors} to the surfaces, $E_3=-I_3I$ and $e_3=-I_3i$.
$E_3$ and $e_3$ are unit vectors perpendicular to the other frame vectors, so
$E^3=E_3$ and $e^3=e_3$.  $\{E_a\}$ and $\{e_a\}$ now both form a basis of
$\mathbb E^3$. The (scalar) {\bf volume forms} $\VOL$ and $\vol$ are defined to
satisfy $\VOL I=E_1\wedge E_2$ and $\vol i=e_1\wedge e_2$.

The {\bf vector derivative} of $\mathbb E^3$ is denoted $\nabla$, and the
projection of this derivative operator onto either $B$ or $S$ is denoted
$\partial$.  $\partial$ can be written locally on $B$ as
$\partial=E^i\frac{\partial}{\partial X^i}$, and on $S$ as
$\partial=e^i\frac{\partial}{\partial x^i}$. For convenience we define the
notation $\partial_i=\frac{\partial}{\partial X^i}$ and
$\partial_i=\frac{\partial}{\partial x^i}$. It will be clear from context
whether differentiation is on the reference or spatial configuration.

Let $\mathsf G(Y)=Y$ and $\mathsf g(y)=y$ be identity functions, where $Y$ is a
vector on $B$, and $y$ is a vector on $S$. The reason we distinguish these
apparently identical linear functions is that they are the {\bf metrics} of the
two surfaces, also called the {\bf first fundamental forms}. In component form
we have $\mathsf g_{ab}=e_a\cdot\mathsf g(e_b)=e_a\cdot e_b$ and $\mathsf
g^{ab}=e^a\cdot e^b$. The properties of the reciprocal frame imply that
$\tensor{\mathsf g}{^a_b}=\tensor{\mathsf g}{_b^a}=\delta^a_b$. Analogous
results hold for $\mathsf G$. The determinant of a function is defined in a
coordinate free way by $\mathsf g(i)=(\det\mathsf g)i$, from which it is clear
that $\det\mathsf G=\det\mathsf g=1$. However, it is common to define
$\det(\mathsf g_{ij})=\mathsf g_{11}\mathsf g_{22}-\mathsf g_{12}\mathsf
g_{21}$, which is not equal to $1$, and in fact encodes important geometric
information about the manifold. This is possible because the coordinate free
definition of $\det\mathsf g$ corresponds to $\det(\tensor{\mathsf g}{^i_j})$,
and not $\det(\mathsf g_{ij})$. In fact, we can show that $\det(\mathsf
g_{ij})=\vol^2$. Recalling the definition of $\vol$, this demonstrates in a very
obvious way that $\sqrt{\det(\mathsf g_{ij})}$ is a measure of the ``volume''
spanned by the parallelepiped formed from the basis vectors. GA in this instance
provides clarification over the fact that $\mathsf g$ is simply the identity
function, and provides a definition of $\vol=\sqrt{\det(\mathsf g_{ij})}$ that
makes its geometric significance immediately obvious.

We denote the {\bf second fundamental forms} on $B$ and $S$ by $\mathsf B$ and
$\mathsf b$. $\mathsf b$ is defined to satisfy $\mathsf
b(y)=-\dot\partial(y\cdot\dot e_3)$. In component form we have $\mathsf
b_{ij}=-e_j\cdot\frac{\partial e_3}{\partial x^i}=e_3\cdot\frac{\partial
  e_j}{\partial x^i}$, which follows from the fact that $e_j\cdot
e_3=0\Rightarrow\partial_i(e_j\cdot e_3)=0$. From this it is clear that $\mathsf
b_{ij}=\mathsf b_{ji}$, and hence $\mathsf b$ is symmetric, i.e. $\mathsf
b(y)=\bar{\mathsf b}(y)=-y\cdot\partial e_3$. The eigenvalues of $\mathsf b$ are
the {\bf principal curvatures} of the surface, denoted by $c_1$ and $c_2$.
Analogous results hold for $\mathsf B$, whose eigenvalues are denoted $C_1$ and
$C_2$. We define the {\bf Christoffel coefficients}
$\gamma^a_{ib}=e^a\cdot\frac{\partial e_b}{\partial x^i}=-e_b\cdot\frac{\partial
  e^a}{\partial x^i}$, which follows from the fact that $e^a\cdot
e_b=\delta^a_b\Rightarrow\partial_i(e^a\cdot e_b)=0$.  $\gamma^i_{jk}$ are the
usual coefficients associated with a frame on a manifold. The remaining
coefficients are closely related to the second fundamental form by
$\gamma^3_{ij}=\mathsf b_{ij}$, $\gamma^i_{j3}=-\tensor{\mathsf b}{^i_j}$, and
$\gamma^3_{i3}=0$, since $e_3$ is a unit vector. We similarly define
$\Gamma^a_{ib}=E^a\cdot\frac{\partial E_b}{\partial X^i}$.

When considering angular momentum we will make use of {\bf bivectors}. To do
this we first introduce some notation. The space of all bivectors in $\mathbb
E^3$ is spanned by the basis $\{e_{(1,3)}=e_1\wedge e_3,e_{(2,3)}=e_2\wedge
e_3,e_{(1,2)}=e_1\wedge e_2\}$, and by the reciprocal basis
$\{e^{(1,3)}=e^3\wedge e^1,e^{(2,3)}=e^3\wedge e^2,e^{(1,2)}=e^2\wedge e^1\}$.
We use capital indices to denote bivector indices, and use the convention that
the indices $I,J,K,\hdots$ run over $(1,3),(2,3)$, while the indices
$A,B,C,\hdots$ run over $(1,3),(2,3),(1,2)$. Hence the space of bivectors is
spanned by $\{e_A\}$ and $\{e^A\}$. Defined in this way these basis bivectors
satisfy $e^A\cdot e_B=\delta^A_B$. In an analogous way to vectors, the general
bivector $\omega$ can be written in component form as
$\omega=\omega_Ae^A=\omega^Ae_A$ where $\omega_A=\omega\cdot e_A$ and
$\omega^A=\omega\cdot e^A$ (here we follow the conventions of
\citep[eq.1-3.18]{Hestenes:1984vg}). We can also define the bivector Christoffel
coefficients $\gamma^A_{iB}=e^A\cdot\frac{\partial e_B}{\partial x^i}$. Given
the surface we have already defined, for which $e_3=e^3$, these satisfy,
\begin{equation}
  \begin{gathered}
    \gamma^{(1,2)}_{i(1,2)} = \gamma^1_{i1} + \gamma^2_{i2},\quad
    \gamma^{(1,3)}_{i(1,2)} = \mathsf b_{i2},\quad
    \gamma^{(2,3)}_{i(1,2)} = -\mathsf b_{i1}, \\
    \gamma^{(1,2)}_{i(1,3)} = -\tensor{\mathsf b}{^2_i},\quad
    \gamma^{(1,3)}_{i(1,3)} = \gamma^1_{i1},\quad
    \gamma^{(2,3)}_{i(1,3)} = \gamma^2_{i1}, \\
    \gamma^{(1,2)}_{i(2,3)} = \tensor{\mathsf b}{^1_i},\quad
    \gamma^{(1,3)}_{i(2,3)} = \gamma^1_{i2},\quad
    \gamma^{(2,3)}_{i(2,3)} = \gamma^2_{i2}.
  \end{gathered}
\end{equation}
Let $\mathsf m$ be a bivector valued function of a vector. $\mathsf m(y)$ can be
written as $\mathsf m(y)=\mathsf m^{Aa}y_a$ where $y_a=y\cdot e_a$ and $\mathsf
m^{Aa}=e^A\cdot\mathsf m(e^a)$, for example, $\mathsf m^{(1,2)2}=(e^2\wedge
e^1)\cdot\mathsf m(e^2)$.

\section{Kinematics}
\label{sec:Kinematics}
We define $X(\eta)$ to be a path over $B$ parametrised by the scalar $\eta$.
$\frac{dX}{d\eta}$ is then a tangent vector to $B$, and we can also obtain a
tangent vector to $S$, $\frac{\partial\phi_t(X)}{\partial\eta}$. The map between
these tangent vectors is denoted $\mathsf F$, and is called the {\bf deformation
  gradient}.  This encodes stretching information for the surface, but also
rigid body rotations. Rigid body rotations are not expected to influence
constitutive theory, so we construct the {\bf Cauchy-Green tensor} $\mathsf
C(Y)=\bar{\mathsf F}\mathsf F(Y)$, which is symmetric. We restrict ourselves to
deformations that have an inverse and leave the orientation of $B$ unchanged,
which means that the eigenvalues of $\mathsf C$ will be real and positive. It is
therefore meaningful to define $\lambda_i$ as the square roots of the
eigenvalues of $\mathsf C$.  These are the {\bf principal stretches} of the
surface. Using the Cauchy-Green tensor we construct the {\bf Green-Lagrange
  strain tensor}, $\mathsf E(Y)=\frac{1}{2}(\mathsf C(Y)-Y)$,that is only
non-zero when the material is locally stretched. Given that $\{x^i\}$ are
convected coordinates, $e_i=\mathsf F(E_i)$. This allows us to obtain the
component expressions $\mathsf C_{ij}=\mathsf F(E_i)\cdot\mathsf F(E_j)=\mathsf
g_{ij}$ and $\mathsf E_{ij}=\frac{1}{2}(\mathsf g_{ij}-\mathsf G_{ij})$. Hence
we see that using convected coordinates, the metric can be used to encode
stretching information. However, our definition is coordinate free.

In $3$-dimensional elasticity the strain tensor is sufficient to characterise
linear constitutive theory. When dealing with shells we must also consider the
bending of the shell, or more precisely, the change of curvature from the
reference to the spatial configuration. Hence we define the {\bf change of
  curvature tensor} $\mathsf H(Y)=\bar{\mathsf F}\mathsf b\mathsf F(Y)-\mathsf
B(Y)$. Using convected coordinates we obtain the component expression $\mathsf
H_{ij}=\mathsf b_{ij}-\mathsf B_{ij}$.

We are also interested in the {\bf strain rate}, and to this end we consider the
rate of change of a tangent vector as it is convected with the surface,
$\frac{\partial\mathsf F(Y)}{\partial t}=\frac{\partial^2\phi_t(X)}{\partial
  t\partial\eta}=\frac{\partial}{\partial\eta}
\frac{\partial\phi_t(X)}{\partial t}=\mathsf F(Y)\cdot\partial v=Y\cdot\partial
V$ where $v$ and $V$ are the velocities referred to the spatial and reference
configurations respectively (see \figref{fig:surfaceGeometry}). Using the fact
that $e_3$ is always normal to $\{e_i\}$ we can write \mbox{$\frac{\partial
    e_3}{\partial t}=-e_3\cdot\frac{\partial e_i}{\partial
    t}e^i=-e_3\cdot(e_i\cdot\partial v)e^i=-v_{3|i}e^i$}, where $v_{a|i}$ is
defined by $v_{a|i}=e_a\cdot(e_i\cdot\partial v)$. Combining these we can now
construct a function that returns the rate of change of a vector, that need not
be tangential to $S$, as it is convected with the motion $\phi_t$. We denote
this function $\mathsf l(y)=\frac{\partial y}{\partial t}=y\cdot\partial
v+y\cdot e_3\frac{\partial e_3}{\partial t}$ (note that $y$ need not be
tangential to $S$ in this expression). It is useful to decompose this into its
symmetric and antisymmetric parts $\mathsf n(y)=\frac{1}{2}(\mathsf
l(y)+\bar{\mathsf l}(y))$ and $\mathsf w(y)=\frac{1}{2}(\mathsf
l(y)-\bar{\mathsf l}(y))$. After some manipulation, the components of $\mathsf
n$ and $\mathsf w$ are given, in terms of convected coordinates, by $\mathsf
n_{ij}=\frac{1}{2}(v_{i|j}+v_{j|i})$, $\mathsf n_{3i}=\mathsf n_{i3}=\mathsf
n_{33}=0$, $\mathsf w_{ij}=\frac{1}{2}(v_{i|j}-v_{j|i})$, $\mathsf
w_{3i}=-\mathsf w_{i3}=v_{3|i}$, and $\mathsf w_{33}=0$.

The symmetric tensor $\mathsf n(y)$ is closely related to $\mathsf E$. We define
the {\bf rate of change of the strain tensor} $\mathsf E$ with time by
$\dot{\mathsf E}(Y)=\frac{\partial\mathsf E(Y)}{\partial t}$. The components of
this tensor are given by $\dot{\mathsf E}_{ij}=\mathsf n_{ij}$, and hence we see
that $\dot{\mathsf E}(Y)=\bar{\mathsf F}\mathsf n\mathsf F(Y)$. This will be
important in constitutive theory.

The {\bf rate of change of the change of curvature tensor} $\dot{\mathsf
  H}(Y)=\frac{\partial\mathsf H(Y)}{\partial t}$ can be expressed in component
form as $\dot{\mathsf H}_{ij}=\frac{\partial\mathsf l_{3j}}{\partial
  x^i}-\gamma^k_{ij}\mathsf l_{3k}-\gamma^a_{i3}\mathsf l_{aj}=\mathsf
l_{3j|i}=e_3\cdot(e_i\cdot\dot\partial\dot{\mathsf l}(e_j))$ (for details see
\S\ref{app:CurvatureRateOfChange}). We see from this that the rate of change of
$\mathsf H$ with time is the $e_3$ component of the second spatial derivative of
velocity. Note that both $\dot{\mathsf E}$ and $\dot{\mathsf H}$ are symmetric.

$\mathsf w$ is an antisymmetric function mapping vectors on $S$ into vectors in
$\mathbb E^3$. Hence, it has a single characteristic eigenbivector $\omega$ such
that $\mathsf w(y)=y\cdot\omega$, which we can extract as
$\omega=\frac{1}{2}e^a\wedge\mathsf w(e_a)$ \citep[\S3.4]{Hestenes:1984vg}.
Defined in this way $\omega$ is the local {\bf angular velocity} of the shell
material, represented as a bivector.  The vector representation of angular
velocity is given by $-I_3\omega$. If we consider $e_3$ being convected with a
material point on the surface, then the fact that it is defined to be a unit
vector allows us to use $\omega$ to write $\frac{\partial e_3}{\partial
  t}=e_3\cdot\omega$. We need a representation of angular velocity in shell
theory since it is not possible to assume, as it is in $3$-dimensional
elasticity, that couple-stresses are negligible. The bivector representation of
angular velocity allows for a much more physical representation of the governing
laws of shells than that suggested by \citep{Naghdi:1972ul}, who requires the
use of a rotated angular velocity with components normal to the shell removed.

\section{Stress}
We consider an arbitrary region of the shell defined by $U\subset B$, which
under the motion $\phi_t$ moves to $\phi_t(U)\subset S$. In continuum mechanics
it is standard to assume that all the forces on the region $U$ can be described
by either body forces or boundary forces, to which we must add body and boundary
moments in shell theory. {\bf Body forces} are expressed in terms of the body
force per unit mass $b(x,t)$. The force acting on the region $U$ due to body
forces is given by,
\begin{equation}
  \int_{\phi_t(U)}\rho b\;\abs{dx} = \int_U\rho b\det\mathsf F\;\abs{dX},
\end{equation}
where $\rho$ is the mass per unit area of the shell, $b(X,t)=b(\phi_t(X),t)$,
and $dx,dX$ are directed volume elements on the spatial and reference
configurations. Directed integration theory is introduced by
\citep[\S6.4]{Doran:2003jd}. In shell theory we must also consider {\bf body
  moments}.  We define the body moment per unit mass $c$ such that the moment
acting on the region $U$ due to body moments is given by,
\begin{equation}
  \int_{\phi_t(U)}\rho c\;\abs{dx} = \int_U\rho c\det\mathsf F\;\abs{dX},
\end{equation}

Next we consider boundary forces and moments. We denote a small portion of the
boundary $\partial\phi_t(U)$ by $\Delta s$, with normal vector $n$. We assume
that the material on the outside of $\Delta s$ exerts a force $\Delta f$, and
moment $\Delta m$ on the material inside. The stress principle of Euler and
Cauchy, adapted for a shell, states,
\begin{quote}
  as the length $\Delta s$ tends to zero, the ratios $\Delta f/\Delta s$ and
  $\Delta m/\Delta s$ tend to definite limits. Moreover, if two paths passing
  through a point $x$ have the same normal $n$, then $\Delta f/\Delta s$ and
  $\Delta m/\Delta s$ tend to the same value for both of these paths
  \citep{Fung:1969ty}.
\end{quote}
Using arguments outlined by, among others, \citep{Fung:1969ty}, we can show that
the limits described in this principle can be expressed as linear functions of
the normal vector $n$ at each point $x\in S$, given a particular time. This
allows us to define the {\bf Cauchy stress tensor} $\sigma(n)$ and the
{\bf couple-stress tensor} $\mathsf m(n)$. We can then write the force on a
portion of the shell due to boundary forces, and the moment on a portion of the
shell due to couple-stresses, as,
\begin{equation}
  \int_{\partial\phi_t(U)}\sigma(n)\abs{ds} =
  \int_{\phi_t(U)}\dot\sigma(\dot\partial)\abs{dx},\quad
  \int_{\partial\phi_t(U)}\mathsf m(n)\abs{ds} =
  \int_{\phi_t(U)}\dot{\mathsf m}(\dot\partial)\abs{dx},
\end{equation}
where $ds$ is a directed boundary element, related to the normal vector by
$n\abs{ds}=ds i^{-1}$.

$\sigma$ and $\mathsf m$ are both tensors on the spatial configuration. We wish
to express balance laws on the reference configuration, so we construct the {\bf
  first Piola-Kirchhoff stress tensor} $\mathsf T$, and the {\bf first reference
  couple-stress tensor} $\mathsf M$, by $\mathsf T(N)=\det\mathsf
F\;\sigma\bar{\mathsf F}^{-1}(N)$ and $\mathsf M(N)=\det\mathsf
F\;\mathsf m\bar{\mathsf F}^{-1}(N)$. Using these we can write,
\begin{equation}
  \int_{\partial\phi_t(U)}\sigma(n)\abs{ds} =
  \int_{\partial U}\mathsf T(N)\abs{dS},\quad
  \int_{\partial\phi_t(U)}\mathsf m(n)\abs{ds} =
  \int_{\partial U}\mathsf M(N)\abs{dS},
\end{equation}
where $dS$ is a directed boundary element on the reference configuration,
related to the normal vector by $N\abs{dS}=dS I^{-1}$. For reasons that become
clearer when considering conservation of energy and constitutive law, we also
define the {\bf second Piola-Kirchhoff stress tensor} $\mathsf S(N)=\mathsf
F^{-1}\mathsf T(N)$ and the {\bf second reference couple-stress tensor} $\mathsf
N(N)=\mathsf F^{-1}\mathsf M(N)$.

The domain of $\sigma$ is vectors tangential to $S$, but its range is $\mathbb
E^3$, and similarly the domains of $\mathsf T$ and $\mathsf S$ are vectors
tangential to $B$, while their ranges are $\mathbb E^3$. $\mathsf m$ is not
vector valued, but bivector valued, since it represents a moment. Its domain is
vectors tangential to $S$, and its range is the space of all bivectors in
$\mathbb E^3$. However, we know that the moments represented by $\mathsf m$ are
due to the stress distribution through the thickness of the shell, and this
means that its range is more restricted. More precisely, we can say that
$\mathsf m^{(1,2)i}=(e^2\wedge e^1)\cdot\mathsf m(e^i)=0$.  Similarly, we assume
that $c$ is due only to shear stresses acting on the upper and lower surfaces of
the shell, meaning that its $e_1\wedge e_2$ component is zero. Given the use of
convected coordinates, the following coordinate expressions hold,
\begin{equation}
  \begin{gathered}
    e^a\cdot\mathsf T(E^i) = \mathsf T^{ai} = E^a\cdot\mathsf S(E^i) =
    \mathsf S^{ai}, \\
    e^I\cdot\mathsf M(E^i) = \mathsf M^{Ii} = E^I\cdot\mathsf N(E^i) =
    \mathsf N^{Ii}.
  \end{gathered}
\end{equation}

It is convenient to define the {\bf modified first reference couple stress
  tensor} \mbox{$\bm{\mathsf M}(N)=\mathsf M(N)\cdot e_3$} and the {\bf modified
  second reference couple stress tensor} \mbox{$\bm{\mathsf N}(N)=\mathsf
  N(N)\cdot E_3$}.  These are vector valued, rather than bivector valued. The
restrictions on the range of $\mathsf m$ imply that the range of $\bm{\mathsf
  M}$ is $TS$, and the range of $\bm{\mathsf N}$ is $TB$. The symmetry of the
$3$-dimensional Cauchy stress tensor implies that $\bm{\mathsf N}$ is symmetric
in the plane $TB$.  These are convenient when expressing conservation of angular
momentum and constitutive laws. Note that $\bm{\mathsf M}$ and $\bm{\mathsf N}$
are not the physical vector representations of the torque. Their natural
emergence in conservation of angular momentum and energy (see
\S\ref{sec:BalanceLaws}) explains why \citep{Naghdi:1972ul} was able to make use
of an apparently unphysical rotated angular velocity in his formulation, and
also justifies the rather strange definition of the vector components of the
couple stress given by \citep[\S1.6.1,eq.1.113]{Leissa:1973tw}. $\bm{\mathsf M}$
and $\bm{\mathsf N}$ satisfy the following coordinate expressions,
\begin{equation}
  \begin{gathered}
    e^1\cdot\bm{\mathsf M}(E^i) = \bm{\mathsf M}^{1i} = \mathsf M^{(1,3)i} =
    \mathsf N^{(1,3)i} = \bm{\mathsf N}^{1i} = E^1\cdot\bm{\mathsf N}(E^i), \\
    e^2\cdot\bm{\mathsf M}(E^i) = \bm{\mathsf M}^{2i} = \mathsf M^{(2,3)i} =
    \mathsf N^{(2,3)i} = \bm{\mathsf N}^{2i} = E^2\cdot\bm{\mathsf N}(E^i).
  \end{gathered}
\end{equation}

\section{Balance Laws}
\label{sec:BalanceLaws}
We write each balance law as an integral equation expressed on the spatial
configuration, and a local equation of motion expressed on the reference
configuration. We are able to express all of these in component free form, which
is a common advantage of using GA.

\begin{description}
  \item[Mass]
    \begin{equation}
      \frac{d}{dt}\int_{\phi_t(U)}\rho\abs{dx} = 0,
    \end{equation}
    \begin{equation}
      \frac{\partial}{\partial t}(\rho\det\mathsf F) = 0.
    \end{equation}
    Using this we can define the time independent density $\rho_0=\rho\det\mathsf
    F$.
  \item[Momentum]
    \begin{equation}
      \frac{d}{dt}\int_{\phi_t(U)}\rho v\abs{dx} =
      \int_{\partial\phi_t(U)}\sigma(n)\abs{ds} +
      \int_{\phi_t(U)}\rho b\abs{dx},
    \end{equation}
    \begin{equation}
      \rho_0\frac{\partial V}{\partial t} = \dot{\mathsf T}(\dot\partial) +
      \rho_0b.
    \end{equation}
  \item[Angular Momentum]
    \begin{equation}
      \frac{d}{dt}\int_{\phi_t(U)}\rho x\wedge v\abs{dx} =
      \int_{\partial\phi_t(U)}x\wedge\sigma(n) + \mathsf m(n)\;\abs{ds} +
      \int_{\phi_t(U)}\rho x\wedge b + \rho c\;\abs{dx},
    \end{equation}
    \begin{equation}
      \dot{\phi_t(X)}\wedge\mathsf T(\dot\partial) +
      \dot{\mathsf M}(\dot\partial) + \rho_0c = 0.
    \end{equation}
    The algebraic manipulations necessary to achieve this expression are given
    in \S\ref{app:ConsOfAngularMomentum}. We can split this expression into its
    $e_1\wedge e_3$, $e_2\wedge e_3$, and $e_1\wedge e_2$ components. These
    components involve taking the divergence of a bivector valued function,
    which is outlined by \citep{Hestenes:1984vg}. However, using the modified
    first couple stress tensor, these components can be written in a more
    familiar form,
    \begin{equation}
      \begin{gathered}
        \mathsf T^{3i} + \frac{\partial\bm{\mathsf M}^{ij}}{\partial X^j} +
        \bm{\mathsf M}^{kj}\gamma^i_{jk} + \bm{\mathsf M}^{ik}\Gamma^j_{jk} +
        \rho c^i = 0, \\
        \mathsf T^{21} - \mathsf T^{12} +
        \bm{\mathsf M}^{2i}\tensor{\mathsf b}{^1_i} -
        \bm{\mathsf M}^{1i}\tensor{\mathsf b}{^2_i} = 0,
      \end{gathered}
    \end{equation}
    where, for convenience, we have defined $c^1=c\cdot(e^3\wedge
    e^1)=c^{(1,3)}$ and $c^2=c\cdot(e^3\wedge e^2)=c^{(2,3)}$. The bivector
    versions of these expressions are equally valid, and easier to interpret
    physically, but less familiar since they involve bivector components. To
    obtain more familiar expressions we need to use modified tensors such as
    $\bm{\mathsf M}$, whose physical meaning is less immediately obvious. The
    use of bivectors to represent angular velocities and torques has illuminated
    why it was necessary for \citep{Naghdi:1972ul} to use apparently unphysical
    quantities to develop his shell theory.

    Conservation of angular momentum has two major implications. The first, from
    the $e_i\wedge e_3$ components of the expression, is that stress normal to
    the tangent plane of the surface are determined if the couple-stress and
    body moment are known. This means that we do not need a constitutive law for
    these components of the stress, we only need constitutive laws for the
    components of stress within the plane of the shell, and for the couple
    stress. The second implication, from the $e_1\wedge e_2$ component, is that
    the {\bf modified second Piola-Kirchhoff stress} $\tilde{\mathsf
      S}(Y)=\mathsf S(Y)-\mathsf F^{-1}\mathsf b\mathsf F\bar{\bm{\mathsf
        N}}(Y)$ is symmetric in \old{$TB$} \new{the tangent space of the
      reference configuration}. This is important when considering conservation
    of energy and in constitutive theory.
  \item[Energy] In this article we assume isothermal elasticity. It is
    uncomplicated to include thermal effects, simply requiring the
    inclusion of the second law of thermodynamics and additional
    constitutive laws, but this extra complication does not contribute to our
    aim here of introducing GA to shell theory for the first time, so is not
    included. Conservation of energy is therefore given by,
    \begin{multline}
      \frac{d}{dt}\int_{\phi_t(U)}\rho\del{e + \frac{v^2}{2}}\abs{dx} =
      \int_{\phi_t(U)}\rho\del{v\cdot b-\omega\cdot c}\abs{dx} \\ +
      \int_{\partial\phi_t(U)} v\cdot\sigma(n)-\omega\cdot\mathsf m(n)\abs{ds},
    \end{multline}
    where $e(x,t)$ is the internal energy per unit mass. The negative signs
    before the moment terms is consistent with the use of bivectors to represent
    moments and angular velocities (see \S\ref{app:BivectorWork}). After some
    algebraic manipulation (see \S\ref{app:ConsOfEnergy}), on the reference
    configuration we obtain,
    \begin{equation}
      \label{eq:consOfEnergy}
      \rho_0\frac{\partial E}{\partial t} = \tr(\tilde{\mathsf S}\dot{\mathsf
        E}) + \tr(\bm{\mathsf N}\dot{\mathsf H}),
    \end{equation}
    where $E(X,t)=e(\phi_t(X),t)$ (which is not the same as $\mathsf E$, the
    Green-Lagrange strain tensor). Note the appearance of the modified second
    Piola-Kirchhoff stress, and the modified second couple stress tensor.  We
    know the first of these is symmetric in $TB$ from conservation of angular
    momentum, and the second must be assumed symmetric in order to obtain a
    determinate theory (this assumption was first proposed by
    \citep[\S15]{Naghdi:1972ul}). This allows us to use this expression to
    derive the constitutive laws given in \S\ref{sec:ConstitutiveTheory}.
\end{description}

\section{Constitutive Theory}
\label{sec:ConstitutiveTheory}
Our basic constitutive assumption is that $E$ is a function of the local values
of the tensors $\mathsf E$ and $\mathsf H$. Applying the chain rule to
\eqref{eq:consOfEnergy}, and noting that the equation is valid for arbitrary
deformations, we obtain the constitutive relations,
\begin{equation}
  \mathsf S(Y) - \mathsf F^{-1}\mathsf b\mathsf F\bm{\mathsf N}(Y) =
  \rho_0\frac{\partial E}{\partial\mathsf E(Y)},\quad
  \bm{\mathsf N}(Y) = \rho_0\frac{\partial E}{\partial\mathsf H(Y)}.
\end{equation}
For an introduction to tensor derivatives, see \citep[\S11.1.2]{Doran:2003jd}.

\citep{Koiter:1966uk} proposes the following form for $\rho_0E$, which can be
regarded as the application of the Saint Venant-Kirchhoff material to shells,
\begin{equation}
  \rho_0E = \frac{E_yh}{2(1-\nu^2)}\del{(1-\nu)\tr(\mathsf E^2) +
    \nu\tr(\mathsf E)^2} + \frac{E_yh^3}{24(1-\nu^2)}\del{
    (1-\nu)\tr(\mathsf H^2) + \nu\tr(\mathsf H)^2},
\end{equation}
where $E_y$ is Young's modulus, $\nu$ is Poisson's ratio, and $h$ is the
thickness of the shell. From this we obtain the following relationships,
\begin{equation}
  \begin{aligned}
    \mathsf S(Y) - \mathsf F^{-1}\mathsf b\mathsf F\bm{\mathsf N}(Y) &=
    \frac{E_yh}{1-\nu^2}\del{(1-\nu)\mathsf E(Y) + \nu\tr(\mathsf E)Y}, \\
    \bm{\mathsf N}(Y) &=
    \frac{E_yh}{12(1-\nu^2)}\del{(1-\nu)\mathsf H(Y) + \nu\tr(\mathsf H)Y}.
  \end{aligned}
\end{equation}
Note that this only provides the part of $\mathsf S$ that is tangential to $B$.
The non-tangential part ($\mathsf S^{3i}$) is found using conservation of
angular momentum.

There is a fundamental contradiction in arriving at the results presented here.
To arrive at the presented form of $\rho_0E$ shown we must make the following
assumptions,
\begin{itemize}
  \item The midsurface in the reference configuration remains the midsurface
    under the motion.
  \item A material line that is normal to the midsurface in the reference
    configuration remains normal to the midsurface.
  \item The shell thickness (measured normal to the midsurface) is constant over
    the surface and does not change with time.
  \item The first and second moments of the density relative to the midsurface
    are zero.
  \item The shell thickness is small compared to its principal radii of
    curvature.
  \item Strains within the shell are small.
  \item Normal stress in the shell is negligible.
\end{itemize}
When applied to Hooke's law in $3$ dimensions, these assumptions imply that the
$e_3$ component of $\sigma$ is zero, but we know that in shell theory these
components are required for conservation of angular momentum. This basic
contradiction remains unresolved.

\section{Linearisation}
We define the displacement $U(X,t)=\phi_t(X)-X$, and we assume that it takes the
form $U=U_0+\epsilon U'$, where $\epsilon$ is small. Neglecting terms of
$\mathcal O(\epsilon^2)$ we obtain,
\begin{equation}
  \begin{aligned}
    \mathsf F(Y) &= Y + Y\cdot\partial U =
    Y + Y\cdot\partial U_0 + \epsilon Y\cdot\partial U' = \mathsf F_0(Y) +
    \epsilon\mathsf F'(Y), \\
    \det\mathsf F &= \det\mathsf F_0 + \epsilon\det\mathsf F_0\tr(\mathsf
    F_0^{-1}\mathsf F'), \\
    \mathsf F^{-1}(Y) &= \mathsf F_0^{-1}(Y) - \epsilon\mathsf F_0^{-1}
    \mathsf F'\mathsf F_0^{-1}(Y), \\
    \det\mathsf F^{-1} &= \det\mathsf F_0^{-1} + \epsilon\det\mathsf
    F_0^{-1}\tr(\mathsf F'\mathsf F_0^{-1}).
  \end{aligned}
\end{equation}
The Green-Lagrange strain tensor can then be written as,
\begin{equation}
  2\mathsf E(Y) = \bar{\mathsf F}_0\mathsf F_0(Y) - Y + \epsilon\del{
    \bar{\mathsf F}_0\mathsf F'(Y) + \bar{\mathsf F}'\mathsf F_0(Y)} =
  2\mathsf E_0(Y) + 2\epsilon\mathsf E'(Y).
\end{equation}
We also need to write the change of curvature tensor in its perturbed form. To
do this we first express the convected basis vectors $\{e_a\}$ as,
\begin{equation}
  \begin{aligned}
    e_i &= \mathsf F_0(E_i) + \epsilon\mathsf F'(E_i), \\
    e_3 &= \frac{\det\mathsf F_0^{-1}}{\VOL}
    \mathsf F_0(E_1)\times\mathsf F_0(E_2) + \epsilon
    \frac{\det\mathsf F_0^{-1}}{\VOL}\left(
    \mathsf F'(E_1)\times\mathsf F_0(E_2) \right. \\
    &\qquad \left. + \mathsf F_0(E_1)\times\mathsf F'(E_2) -
    \tr(\mathsf F'\mathsf F_0^{-1})
    \mathsf F_0(E_1)\times\mathsf F_0(E_2)\right) \\
    &= e_{30} + \epsilon e_3'.
  \end{aligned}
\end{equation}
where $\times$ is the vector cross product, defined by $a\times b=-I_3a\wedge
b$. This allows us to write $\bar{\mathsf F}\mathsf b\mathsf F$ as,
\begin{equation}
  \begin{aligned}
    \bar{\mathsf F}\mathsf b\mathsf F(Y) &= Y^iE^je_{30}\cdot\partial_i
    \mathsf F_0(E_j) + \epsilon Y^iE^j\del{e_{30}\cdot\partial_i\mathsf F'(E_j)
      + e_3'\cdot\partial_i\mathsf F_0(E_j)} \\
    &= (\bar{\mathsf F}\mathsf b\mathsf F)_0(Y) +
    \epsilon(\bar{\mathsf F}\mathsf b\mathsf F)'(Y),
  \end{aligned}
\end{equation}
which in turn allows us to express $\mathsf H$ as,
\begin{equation}
  \mathsf H(Y) = (\bar{\mathsf F}\mathsf b\mathsf F)_0(Y) - \mathsf B(y) +
  \epsilon(\bar{\mathsf F}\mathsf b\mathsf F)'(Y) =
  \mathsf H_0(Y) + \epsilon\mathsf H'(Y).
\end{equation}

We can now write the modified second reference couple stress tensor $\bm{\mathsf
  N}$ as,
\begin{equation}
  \begin{aligned}
    \bm{\mathsf N}(y) &= \frac{Eh^3}{12(1-\nu^2)}\del{(1-\nu)\mathsf H_0(y) +
      \nu\tr(\mathsf H_0)y} + \epsilon\frac{Eh^3}{12(1-\nu^2)}\del{
      (1-\nu)\mathsf H'(y) + \nu\tr(\mathsf H')y} \\
    &= \bm{\mathsf N}_0(y) + \epsilon\bm{\mathsf N}'(y).
  \end{aligned}
\end{equation}
To express the second Piola-Kirchhoff stress tensor we first need to express
$\mathsf F^{-1}\mathsf b\mathsf F(Y)$,
\begin{equation}
  \begin{aligned}
    \mathsf F^{-1}\mathsf b\mathsf F(y) &=
    Y^iE_je_{30}\cdot\partial_i\bar{\mathsf F}_0^{-1}(E^j) +
    \epsilon Y^iE_j
    \del{-e_{30}\cdot\partial_i\bar{\mathsf F}_0^{-1}\bar{\mathsf F}'
      \bar{\mathsf F}_0^{-1}(E^j) + e_3'\cdot\partial_i
      \bar{\mathsf F}_0^{-1}(E^j)} \\
    &= (\mathsf F^{-1}\mathsf b\mathsf F)_0(y) + \epsilon
    (\mathsf F^{-1}\mathsf b\mathsf F)'(y).
  \end{aligned}
\end{equation}
This allows us to write $\mathsf S$ and $\mathsf T$ as,
\begin{equation}
  \begin{aligned}
    \mathsf S(Y) &\approx \frac{Eh}{1-\nu^2}\del{(1-\nu)\mathsf E_0(Y) +
      \nu\tr(\mathsf E_0)Y} + \epsilon\frac{Eh}{1-\nu^2}
    \del{(1-\nu)\mathsf E'(Y) + \nu\tr(\mathsf E')Y} \\
    &\quad + (\mathsf F^{-1}\mathsf b\mathsf F)_0\bm{\mathsf N}_0(Y) +
    \epsilon\del{(\mathsf F^{-1}\mathsf b\mathsf F)_0\bm{\mathsf N}'(Y) +
      (\mathsf F^{-1}\mathsf b\mathsf F)'\bm{\mathsf N}_0(Y)} \\
    &= \mathsf S_0(Y) + \epsilon\mathsf S'(Y), \\
    \mathsf T(Y) &= \mathsf F_0\mathsf S_0(y) + \epsilon\del{
      \mathsf F_0\mathsf S'(y) + \mathsf F'\mathsf S_0(y)} \\
    &= \mathsf T_0(y) + \epsilon\mathsf T'(y).
  \end{aligned}
\end{equation}

Conservation of mass can be expressed as,
\begin{equation}
  \frac{\partial}{\partial t}(\rho\det\mathsf F) =
  \frac{\partial}{\partial t}\del{\rho\del{\det\mathsf F_0 +
      \epsilon\det\mathsf F_0\tr(\mathsf F_0^{-1}\mathsf F')}} = 0.
\end{equation}
We define $\rho_0=\rho\det\mathsf F_0$ and $\rho'=\rho\det\mathsf F_0\tr(\mathsf
F_0^{-1}\mathsf F')$ (note the adjustment of the definition of $\rho_0$). We
assume that the initial displacement $U_0$ satisfies the governing equations
separately, so both $\rho_0$ and $\rho'$ are independent of time. Conservation
of momentum can be expressed as,
\begin{equation}
  (\rho_0+\epsilon\rho')\frac{\partial^2}{\partial t^2}(U_0 + \epsilon U') =
  \dot{\mathsf T}_0(\dot\partial) + \epsilon\dot{\mathsf T}'(\dot\partial) +
  (\rho_0+\epsilon\rho')b.
\end{equation}
We denote the body force acting on the body in its initial deformed state
(defined by $U_0$) by $b_0$, and then decompose $b$ as $b=b_0+\epsilon b'$. This
includes the assumption that the additional body force acting on the body after
the perturbation $\epsilon U'$ is small. Subtracting conservation of momentum
for the initial deformation $U_0$, we obtain,
\begin{equation}
  \rho'\frac{\partial^2U_0}{\partial t^2} +
  \rho_0\frac{\partial^2U'}{\partial t^2} = \dot{\mathsf T}'(\dot\partial) +
  \rho_0b' + \rho'b_0.
\end{equation}
Usually we assume that $U_0$ is time independent, meaning that we obtain,
\begin{equation}
  \rho_0\frac{\partial^2U'}{\partial t^2} = \dot{\mathsf T}'(\dot\partial) +
  \rho_0b' + \rho'b_0.
\end{equation}

We can write $\mathsf M$ as,
\begin{equation}
  \begin{aligned}
    \mathsf M(Y) &= \mathsf F\bm{\mathsf N}(Y)\wedge e_3 \\
    &= \mathsf F_0\bm{\mathsf N}_0(Y)\wedge e_{30} + \epsilon\del{
      \mathsf F'\bm{\mathsf N}_0(Y)\wedge e_{30} +
      \mathsf F_0\bm{\mathsf N}'(Y)\wedge e_{30} +
      \mathsf F_0\bm{\mathsf N}_0(Y)\wedge e_3'} \\
    &= \mathsf M_0(y) + \epsilon\mathsf M'(y).
  \end{aligned}
\end{equation}
If, as with $b$, we assume that $c$ can be decomposed as $c=c_0+\epsilon c'$,
then this allows us to write the perturbed part of conservation of angular
momentum as,
\begin{equation}
  \mathsf F_0(E_i)\wedge\mathsf T'(E^i) +
  \mathsf F'(E_i)\wedge\mathsf T_0(E^i) +
  \dot{\mathsf M}'(\dot\partial) + \rho_0c' + \rho'c_0 = 0.
\end{equation}

\subsection{Small Displacements}
If we assume $U_0=0$ (or that it is constant) then we obtain the following
simplifications,
\begin{equation}
  \begin{aligned}
    \mathsf F(Y) &= Y + \epsilon Y\cdot\partial U' = U +
    \epsilon\mathsf F'(Y), \\
    \mathsf F^{-1}(Y) &= Y - \epsilon\mathsf F'(Y), \\
    2\mathsf E(y) &= \epsilon\del{\mathsf F'(Y) +
      \bar{\mathsf F}'(Y)} = Y\cdot\partial U + \dot\partial(Y\cdot\dot U), \\
    \det\mathsf F &= 1 + \epsilon\tr(\mathsf F') =
    1 + \epsilon\partial\cdot U', \\
    \det\mathsf F^{-1} &= 1 - \epsilon\tr(\mathsf F') =
    1 - \epsilon\partial\cdot U', \\
    e_i &= E_i + \epsilon\mathsf f'(E_i) =
    E_i + \epsilon E_i\cdot\partial U', \\
    e^i &= E^i - \epsilon\bar{\mathsf F}'(E^i) = E^i -
    \epsilon\dot\partial(E^i\cdot\dot U'), \\
    e_3 &= E_3 + \epsilon\del{
      \frac{1}{\VOL}(E_1\cdot\partial U')\times E_2 +
      \frac{1}{\VOL}E_1\times(E_2\cdot\partial U') -
      (\partial\cdot u')E_3}. \\
  \end{aligned}
\end{equation}
Following the method of \citep{CiarletJr:2005vn}, and using the coordinate
independent notation of GA, the change of curvature tensor takes the form,
\begin{equation}
  \mathsf H(Y) = \epsilon E^j\del{(E_j\cdot
    \dot\partial\dot{\mathsf F}'(Y))\cdot E_3}.
\end{equation}
From this it is clear how the linearised change of curvature tensor is closely
related to the $E_3$ components of the second derivative of the displacement
field $U'$.

Using these results and applying the formulas of the previous section we have,
$\mathsf S_0(Y)=0$, $\mathsf T_0(Y)=0$, $\bm{\mathsf N}_0(Y)=0$, $\mathsf
M_0(Y)=0$, and,
\begin{equation}
  \begin{aligned}
    \bm{\mathsf N}'(Y) &= \frac{E_yh^3}{12(1-\nu^2)}\del{
      (1-\nu)\mathsf H'(Y) + \nu\tr(\mathsf H')Y}. \\
    \mathsf S(Y) &= \frac{E_yh}{1-\nu^2}\del{
      (1-\nu)\mathsf E'(Y) + \nu\tr(\mathsf E')Y} \\
    &\qquad + \frac{E_yh^3}{12(1-\nu^2)}\del{
      (1-\nu)\mathsf B\mathsf H'(Y) + \nu\tr(\mathsf H')\mathsf B(Y)}, \\
    \mathsf M'(Y) &= \bm{\mathsf N}'(Y)\wedge E_3.
  \end{aligned}
\end{equation}
Conservation of momentum and angular momentum can be written as,
\begin{gather}
  \rho_0\frac{\partial^2U'}{\partial t^2} =
  \dot{\mathsf S}'(\dot\partial) + \rho_0b', \\
  E_i\wedge\mathsf S'(E^i) + \dot{\mathsf M}'(\dot\partial) + \rho' c_0 = 0.
\end{gather}

It is worth pointing out an anomaly at this point, which is made clearer by the
use of geometric algebra. Much of the previous work done on linearised shell
theory (summarised by \citep{Leissa:1973tw}, with one of the more rigorous
derivations provided by \citep{Vlasov:1951tg,Vlasov:1964wc}) uses a rather
strange coordinate system, which does not aid comprehension. Coordinates are
chosen such that the lines $X^i=constant$ define lines of principal curvature on
the reference configuration. This allows the components of several tensors to be
expressed more simply, since the basis vectors are orthogonal, and are
eigenvectors of $\mathsf B$. However, the coordinate system is not constrained
to be \emph{orthonormal}, meaning that the reciprocal frame and frame do not
coincide (though $E^i$ is parallel to $E_i$). This is not a problem in geometric
algebra, since we can use an arbitrary coordinate system, but the solution
adopted by many authors is to create a new normalised frame $\cbr{\hat
  E_i=\frac{E_i}{\abs{E_i}}}$. Differentiation is performed with respect to the
coordinates $\{X^i\}$, but tensor and vector components are expressed relative to
the frame $\{\hat E_i\}$. This adds considerable complication to the expressions
for strain and change of curvature, which, through the use of geometric algebra,
we have simplified.

\subsection{Uni-Axial Strain of a Cylinder}

We now consider the case in which $\{E_i\}$ form an orthonormal basis, and are
also the eigenvectors of $\mathsf B$. In this case we have $\Gamma^i_{jk}=0$
and,
\begin{equation}
  \mathsf B_{11} = \Gamma^3_{11} = -\Gamma^1_{13} = C_1,\quad
  \mathsf B_{12} = \Gamma^3_{12} = -\Gamma^2_{13} = 0,\quad
  \mathsf B_{22} = \Gamma^3_{22} = -\Gamma^2_{23} = C_2.
\end{equation}
Moreover, we take $C_1=0$ and $C_2=C$. We take the background deformation to be
uni-axial strain such that $U_0=\varepsilon X^1E_1$. $X^1$ is the axial distance
along the cylindrical shell, and $X^2$ is the azimuthal distance around the
circumference. We can write $\mathsf F_0$ as $\mathsf F_0(Y)=Y+\varepsilon
(Y\cdot E_1)E_1$, but $E_1$ is a basis vector specific to the tangent space of
the reference configuration. For clarity we therefore define a unit vector
aligned with the axis of symmetry of the cylindrical shell $\bm e$, which is
defined everywhere in $\mathbb E^3$. On the reference configuration $\bm e=E_1$.
Using this we write,
\begin{equation}
  \begin{aligned}
    \mathsf F_0(Y) &= Y + \varepsilon(Y\cdot\bm e)\bm e, \\
    \bar{\mathsf F}_0(y) &= y + \varepsilon(y\cdot\bm e)\bm e, \\
    \mathsf f_0^{-1}(y) &= y -
    \tfrac{\varepsilon}{\lambda}(y\cdot\bm e)\bm e, \\
    \bar{\mathsf F}_0^{-1}(Y) &= Y -
    \tfrac{\varepsilon}{\lambda}(Y\cdot\bm e)\bm e, \\
    \det\mathsf F_0 &= 1+\varepsilon = \lambda,
  \end{aligned}
\end{equation}
where we have also defined $\lambda=1+\varepsilon$. Given that the frame $E_i$
is orthonormal, we do not need to distinguish sub- and superscript indices.
Hence we can obtain the following expression for the components of the tensors
derived in previous sections,
\begin{equation}
  \begin{gathered}
    (\mathsf E_0)_{11} = \tfrac{1}{2}\varepsilon^2+\varepsilon,\quad
    (\mathsf E_0)_{12} = (\mathsf E_0)_{21} = 0,\quad
    (\mathsf E_0)_{22} = 0, \\
    (\mathsf E')_{11} = (1+\varepsilon)\partial_1U'_1,\quad
    (\mathsf E')_{22} = \partial_2U'_2 - CU'_3, \\
    (\mathsf E')_{12} = (\mathsf E')_{21} =
    \tfrac{1}{2}(\partial_1U'_2 + \partial_2U'_1) +
    \tfrac{\varepsilon}{2}\partial_2U'_1.
  \end{gathered}
\end{equation}
\begin{equation}
  \begin{gathered}
    (\mathsf H_0)_{ij} = 0, \\
    \mathsf H'_{11} = \partial_1\partial_1U'_3, \quad
    \mathsf H'_{22} = \partial_2\partial_2U'_3 + 2C\partial_2U'_2 - C^2U'_3, \\
    \mathsf H'_{12} = \mathsf H'_{21} = \partial_1\partial_2U'_3 +
    C\partial_1U'_2.
  \end{gathered}
\end{equation}
\begin{equation}
  \begin{gathered}
    (\bm{\mathsf N}_0)_{ij} = 0, \\
    \bm{\mathsf N}'_{11} = \frac{E_yh^3}{12(1-\nu^2)}(\mathsf H'_{11} +
    \nu\mathsf H'_{22}),\quad
    \bm{\mathsf N}'_{22} = \frac{E_yh^3}{12(1-\nu^2)}(\mathsf H'_{22} +
    \nu\mathsf H'_{11}), \\
    \bm{\mathsf N}'_{12} = \bm{\mathsf N}'_{21} =
    \frac{E_yh^3}{12(1+\nu)}\mathsf H'_{12}.
  \end{gathered}
\end{equation}
\begin{equation}
  \begin{gathered}
    (\mathsf S_0)_{11} = \frac{E_yh}{1-\nu^2}(\mathsf E_0)_{11},\quad
    (\mathsf S_0)_{22} = \frac{E_yh}{1-\nu^2}\nu(\mathsf E_0)_{11}, \\
    (\mathsf S_0)_{12} = (\mathsf S_0)_{21} = 0.
  \end{gathered}
\end{equation}
\begin{equation}
  \begin{gathered}
    \mathsf S'_{11} = \frac{E_yh}{1-\nu^2}(\mathsf E'_{11} +
    \nu\mathsf E'_{22}) + C\bm{\mathsf N}'_{22},\quad
    \mathsf S'_{22} = \frac{E_yh}{1-\nu^2}(\mathsf E'_{22} +
    \nu\mathsf E'_{11}), \\
    \mathsf S'_{12} = \frac{E_yh}{1+\nu}\mathsf E'_{12},\quad
    \mathsf S'_{21} = \frac{E_yh}{1+\nu}\mathsf E'_{21} +
    C\bm{\mathsf N}'_{21}.
  \end{gathered}
\end{equation}
This demonstrates the application of linearised shell theory to a situation
where there is prior strain.

\section{Conclusions}
The elastic theory of shells has been advanced using geometric algebra,
providing clarifications and some new developments. We have provided a lucid,
geometric interpretation of $\det(\mathsf G_{ij})=\VOL^2$, and clarified the
difference between the coordinate definition $\det(\mathsf G_{ij})$ and the
coordinate free definition of the determinant of $\mathsf G$. As has been
the case in other areas, geometric algebra has allowed a coordinate free
representation of balance laws and constitutive laws, which makes physical
interpretation clearer, while also providing the tools to easily express these
equations in terms of arbitrary coordinate systems for practical purposes. The
role of moments and angular velocity, and the apparent use by previous authors
of an unphysical angular velocity, has been clarified through the use of a
bivector representation. \new{We hope that this early work using GA will allow
  the powerful encoding of rotations by GA, using rotors, to be used in a
  similar way as has been done for rods \citep{McRobie:1999wv}.} In linearised
theory clarification of confusing previous coordinate conventions has been
provided, and the introduction of prior strain into the linearised theory of
shells has been made possible.

\vskip6pt

\enlargethispage{20pt}

\paragraph{Ethics and permission to carry out field work.} Not applicable.

\paragraph{Data accessibility.} No data is associated with this paper.

\paragraph{Authors' contributions.} The majority of the work was completed by
AG, with JL and AA providing advice both at early conceptual stages and during
drafting.

\new{\paragraph{Acknowledgements.} The friendly and accommodating staff of the
  coffee shop Hot Numbers, across the road from the Engineering Department, who
  provided a place for many useful discussions.}

\paragraph{Competing interests.} The authors declare no competing interests.

\paragraph{Funding.} The authors would like to acknowledge funding from the
EPSRC, the IMechE Postgraduate Research Scholarship, and Engineering for
Clinical Practice (http://divf.eng.cam.ac.uk/ecp/Main/EcpResearch).

\appendix
\renewcommand{\theequation}{\Alph{section}.\arabic{equation}}

\section{The Rate of Change of Curvature}
\label{app:CurvatureRateOfChange}
The components of $\dot{\mathsf H}$ are given by $\dot{\mathsf
  H}_{ij}=\frac{\partial\mathsf H_{ij}}{\partial t}$. The component expression
$\mathsf H_{ij}=\mathsf b_{ij}-\mathsf B_{ij}$ has already been derived, and
$\mathsf B$ does not change with time, so we can write,
\begin{equation}
  \begin{aligned}
    \dot{\mathsf H}_{ij} = \frac{\partial\mathsf H_{ij}}{\partial t} =
    \frac{\partial\mathsf b_{ij}}{\partial t} &= \frac{\partial}{\partial t}
    \del{e_3\cdot\frac{\partial e_j}{\partial x^i}} \\
    &= \frac{\partial e_3}{\partial t}\cdot\frac{\partial e_j}{\partial x^i} +
    e_3\cdot\frac{\partial}{\partial x^i}\frac{\partial e_j}{\partial t} \\
    &= \mathsf l_{a3}\gamma^a_{ij} + e_3\cdot\partial_i\del{\mathsf l_{aj}e^a}
    \\
    &= -\mathsf l_{3a}\gamma^a_{ij} + \partial_i\mathsf l_{3j} -
    \mathsf l_{aj}\gamma^a_{i3} \\
    &= \partial_i\mathsf l_{3j} - \gamma^k_{ij}\mathsf l_{3k} -
    \gamma^a_{i3}\mathsf l_{aj},
  \end{aligned}
\end{equation}
Here we have used the fact that $\mathsf l_{3i}=-\mathsf l_{i3}$, and $\mathsf
l_{33}=0$. Also, where appropriate, time derivatives have been taken assuming
that we are being convected with the surface. The spatial derivative of a
function defined on the surface $S$ is given by,
\begin{equation}
  \begin{aligned}
    y\cdot\dot\partial\dot{\mathsf l}(z) &= y\cdot\partial\del{\mathsf l(z)} -
    \mathsf l\del{y\cdot\partial z} \\
    &= y^i\partial_i(\mathsf l_{aj}z^je^a) -
    \mathsf l\del{y^i\partial_i(z^je_j)} \\
    &= y^i\del{\partial_i(\mathsf l_{aj})z^je^a +
      \mathsf l_{aj}\partial_i(z^j)e^a + \mathsf l_{aj}z^j\partial_i(e^a) -
      \partial_i(z^j)\mathsf l(e_j) - z^j\mathsf l\del{\partial_i(e_j)}} \\
    &= y^iz^j\del{\partial_i(\mathsf l_{aj})e^a -
      \mathsf l_{aj}\gamma^a_{ib}e^b - \gamma^a_{ij}\mathsf l_{ba}e^b} \\
    &= y^iz^j\del{\partial_i\mathsf l_{aj} - \gamma^k_{ij}\mathsf l_{ak} -
      \gamma^b_{ia}\mathsf l_{bj}}e^a = y^iz^je^a\mathsf l_{aj|i},
  \end{aligned}
\end{equation}
where we have assumed that $\mathsf l(e_3)=0$, and the last equality defines
$\mathsf l_{aj|i}$. Hence, we see that $\dot{\mathsf H}_{ij}$ is given by,
\begin{equation}
  \dot{\mathsf H}_{ij} = \partial_i\mathsf l_{3j} -
  \gamma^k_{ij}\mathsf l_{3k} - \gamma^a_{i3}\mathsf l_{aj} =
  \mathsf l_{3j|i} = e_3\cdot(e_i\cdot\dot\partial\dot{\mathsf l}(e_j)).
\end{equation}

\section{Work Done by Bivector Torque}
\label{app:BivectorWork}
Let $\omega$ and $\omega_v$ be the bivector and vector representations of the
angular velocity of a body, related by $\omega=I_3\omega_v$. If $q_v$ is the
vector representation of the torque acting on the body then the rate at which
work is done on the body is given by $\omega_v\cdot q_v$. Making use of
\citep[\S4.1.3]{Doran:2003jd}, we can write this as,
\begin{equation}
  \begin{aligned}
  \omega_v\cdot q_v &= -I_3^2\omega_v\cdot q_v  \\
  &= -\tfrac{1}{2}I_3^2(\omega_vq_v+q_v\omega_v) \\
  &= -\tfrac{1}{2}I_3(I_3\omega_vq_v + I_3q_v\omega_v) \\
  &= -\tfrac{1}{2}I_3(\omega_vI_3q_v + q_vI_3\omega_v) \\
  &= -\tfrac{1}{2}\del{(I_3\omega_v)(I_3q_v) + (I_3q_v)(I_3\omega_v)} \\
  &= -(I_3\omega_v)\cdot(I_3q_v) \\
  &= -\omega\cdot(I_3 q_v)
  \end{aligned}
\end{equation}
The bivector representation of torque $q$ is related to $q_v$ by $q=I_3q_v$ (see
\citep[\S3.1.1]{Doran:2003jd}), and so the rate of work done by the torque $q$
is given by $-\omega\cdot q$.

\section{Conservation of Angular Momentum}
\label{app:ConsOfAngularMomentum}
We can express conservation of angular momentum on the reference configuration
as,
\begin{equation}
  \begin{aligned}
    \frac{d}{dt}\int_U \rho\phi_t(X)\wedge V\;\det\mathsf F\;\abs{dX} &=
    \int_{\partial U}\phi_t(X)\wedge\mathsf T(N) + \mathsf M(N)\;\abs{dS} \\
    &\quad +
    \int_U\del{\rho\phi_t(X)\wedge b + \rho c}\det\mathsf F\;\abs{dX}.
  \end{aligned}
\end{equation}
Simplifying using conservation of mass we can express this in local form as,
\begin{equation}
  \rho_0\phi_t(X)\wedge\frac{\partial V}{\partial t} =
  \phi_t(X)\wedge\dot{\mathsf T}(\dot\partial) +
  \dot{\phi_t(X)}\wedge\mathsf T(\dot\partial) +
  \dot{\mathsf M}(\dot\partial) +
  \rho_0\phi_t(X)\wedge b + \rho_0c.
\end{equation}
Using conservation of momentum this simplifies to,
\begin{equation}
  \dot{\phi_t(X)}\wedge\mathsf T(\dot\partial) +
  \dot{\mathsf M}(\dot\partial) + \rho_0c = 0.
\end{equation}
The first term in this expression can be written as,
\begin{equation}
  \dot{\phi_t(X)}\wedge\mathsf T(\dot\partial) =
  (E_i\cdot\partial\phi_t(X))\wedge\mathsf T(E^i) =
  \mathsf F(E_i)\wedge\mathsf T(E^i) =
  \mathsf F(E_i)\wedge\mathsf F\mathsf S(E^i).
\end{equation}
We can express $\dot{\mathsf M}(\dot\partial)$ as,
\begin{equation}
  \begin{aligned}
    \dot{\mathsf M}(\dot\partial) &= \del{
      \frac{\partial\mathsf M^{Ii}}{\partial X^i} +
      \mathsf M^{Ji}\gamma^I_{iJ} + \mathsf M^{Ij}\Gamma^i_{ij}}e_I +
    \mathsf M^{Ji}\gamma^{(1,2)}_{iJ}e_1\wedge e_2 \\
    &= \tensor{\mathsf M}{^{Ii}_{|i}}e_I +
    \mathsf M^{Ji}\gamma^{(1,2)}_{iJ}e_1\wedge e_2.
  \end{aligned}
\end{equation}
Writing conservation of angular momentum in component form (noting that we must
use the convected frame $\{e_i\}$) we obtain,
\begin{equation}
  \begin{gathered}
    \mathsf S^{31} + \tensor{\mathsf M}{^{(1,3)i}_{|i}} +
    \rho_0c^{(1,3)} = 0, \\
    \mathsf S^{32} + \tensor{\mathsf M}{^{(2,3)i}_{|i}} +
    \rho_0c^{(2,3)} = 0, \\
    \mathsf S^{21} - \mathsf S^{12} +
    \mathsf M^{(2,3)i}\tensor{\mathsf b}{^1_i} -
    \mathsf M^{(1,3)i}\tensor{\mathsf b}{^2_i} = 0.
  \end{gathered}
\end{equation}
By using the modified first reference couple-stress tensor $\bm{\mathsf
  M}(y)=\mathsf M(y)\cdot e_3$ we can write the last of these as,
\begin{equation}
  \mathsf S^{21} - \mathsf S^{12} +
  \bm{\mathsf M}^{2i}\tensor{\mathsf b}{^1_i} -
  \bm{\mathsf M}^{1i}\tensor{\mathsf b}{^2_i} = 0.
\end{equation}
Alternatively, we can use the modified second reference couple stress tensor
$\bm{\mathsf N}$ to write this as,
\begin{equation}
  \mathsf S^{21} - \mathsf S^{12} +
  \bm{\mathsf N}^{2i}\tensor{\mathsf b}{^1_i} -
  \bm{\mathsf N}^{1i}\tensor{\mathsf b}{^2_i} = 0.
\end{equation}
We define the {\bf modified second Piola-Kirchhoff} stress tensor by,
\begin{equation}
  \begin{gathered}
    \tilde{\mathsf S}(y) = \mathsf S(y) -
    \mathsf F^{-1}\mathsf b\mathsf F\bar{\bm{\mathsf N}}(y), \\
    \tilde{\mathsf S}^{ij} = \mathsf S^{ij} -
    \tensor{\mathsf b}{^i_k}\bar{\bm{\mathsf N}}^{kj} =
    \mathsf S^{ij} - \tensor{\mathsf b}{^i_k}\bm{\mathsf N}^{jk},\quad
    \tilde{\mathsf S}^{3i} = \mathsf S^{3i}.
  \end{gathered}
\end{equation}
Conservation of angular momentum then implies that, $\tilde{\mathsf S}^{21} =
\tilde{\mathsf S}^{12}$, hence, $\tilde{\mathsf S}$ is symmetric in the plane of
the shell.

\section{Conservation of Energy}
\label{app:ConsOfEnergy}
Conservation of energy can be expressed on the reference configuration as,
\begin{multline}
  \frac{d}{dt}\int_U\rho\del{E+\frac{V^2}{2}}\det\mathsf F\;\abs{dX} =
  \int_U\rho\del{V\cdot b - \Omega\cdot c}\det\mathsf F\;\abs{dX} \\ +
  \int_{\partial U}V\cdot\mathsf T(n) - \Omega\cdot\mathsf M(n)\;\abs{dS},
\end{multline}
where we have defined $\Omega$ to be the angular velocity referred to the
reference configurations $\Omega(X,t)=\omega(\phi_t(X),t)$. Converting to local
form, and making use of conservation of mass and momentum, this can be written
as,
\begin{equation}
  \rho_0\frac{\partial E}{\partial t} = - \rho_0\Omega\cdot c +
  \dot V\cdot\mathsf T(\dot\partial) - \dot\Omega\cdot\mathsf M(\dot\partial) -
  \Omega\cdot\dot{\mathsf M}(\dot\partial).
\end{equation}
Making use of conservation of angular momentum we obtain,
\begin{equation}
  \rho_0\frac{\partial E}{\partial t} = \dot V\cdot\mathsf T(\dot\partial) -
  \dot\Omega\cdot\mathsf M(\dot\partial) +
  \Omega\cdot(\dot{\phi_t(X)}\wedge\mathsf T(\dot\partial)).
\end{equation}
$\mathsf S$ is related to $\mathsf T$ by $\mathsf T(y)=\mathsf F\mathsf S(y)$,
so we can write conservation of energy as,
\begin{equation}
  \rho_0\frac{\partial E}{\partial t} =
  \dot V\cdot\mathsf F\mathsf S(\dot\partial) -
  \dot\Omega\cdot\mathsf M(\dot\partial) +
  \Omega\cdot(\mathsf F(E_i)\wedge\mathsf F\mathsf S(E^i)).
\end{equation}
The first term on the right hand side in this expression can be expressed as,
\begin{equation}
  (E_i\cdot\partial V)\cdot\mathsf F\mathsf S(E^i) =
  (E_i\cdot\partial V)\cdot\mathsf F(E_a)\mathsf S^{ai} =
  (\mathsf F(E_i)\cdot\partial v)\cdot\mathsf F(E_a)\mathsf S^{ai} =
  v_{a|i}\mathsf S^{ai}.
\end{equation}
We can write $\Omega\cdot(\mathsf F(E_i)\wedge\mathsf F\mathsf
S(E^i))-\dot\Omega\cdot\mathsf M(\dot\partial)$ as,
\begin{equation}
  \begin{aligned}
    \Omega\cdot(\mathsf F(E_i)\wedge\mathsf F\mathsf S(E^i)) -
    \dot\Omega\cdot\mathsf M(\dot\partial) &=
    \omega_Ae^A\cdot(e_i\wedge e_a)\mathsf S^{ai} -
    (e_i\cdot\partial\omega)\cdot e_I\mathsf M^{Ii} \\
    &= \omega_i\mathsf S^{3i} + \omega_3(\mathsf S^{21}-\mathsf S^{12}) \\
    &\quad -
    \del{\frac{\partial\omega_I}{\partial x^i} - \omega_J\gamma^J_{iI} -
      \omega_{(1,2)}\gamma^{(1,2)}_{iI}}\mathsf M^{Ii} \\
    &= \omega_i\mathsf S^{3i} + \omega_3\del{\mathsf S^{21}-\mathsf S^{12} +
      \mathsf M^{(2,3)i}\tensor{\mathsf b}{^1_i} -
      \mathsf M^{(1,3)i}\tensor{\mathsf b}{^2_i}} \\
    &\quad - \del{\frac{\partial\omega_I}{\partial x^i} -
      \omega_J\gamma^J_{iI}}\mathsf M^{Ii} \\
    &= \omega_i\mathsf S^{3i} - \del{\frac{\partial\omega_I}{\partial x^i} -
      \omega_J\gamma^J_{iI}}\mathsf M^{Ii}.
  \end{aligned}
\end{equation}
where, for convenience, we have defined $\omega_1=\omega_{(1,3)}$,
$\omega_2=\omega_{(2,3)}$, and $\omega_3=\omega_{(1,2)}$. The components of
$\omega$ can be found using $\omega=\frac{1}{2}e^a\wedge\mathsf w(e_a)$. The
first two components are given by,
\begin{equation}
  \omega_{(1,3)} = v_{3|1},\quad \omega_{(2,3)} = v_{3|2}.
\end{equation}
Using these expressions for the components of $\omega$ we obtain,
\begin{equation}
  \begin{aligned}
    \Omega\cdot(\mathsf F(E_i)\wedge\mathsf F\mathsf S(E^i)) -
    \dot\Omega\cdot\mathsf M(\dot\partial) &= -v_{3|i}\mathsf s^{3i} +
    \del{\frac{\partial v_{3|i}}{\partial x^j} -
      v_{3|k}\gamma^k_{ji} +
      v_{k|i}\tensor{\mathsf b}{^k_j}}\bm{\mathsf M}^{ij} \\
    &\quad - v_{k|i}\tensor{\mathsf b}{^k_j}\bm{\mathsf M}^{ij} \\
    &= -v_{3|i}\mathsf S^{3i} + \mathsf l_{3i|j}\bm{\mathsf M}^{ij} -
    v_{k|i}\tensor{\mathsf b}{^k_j}\bm{\mathsf M}^{ij}.
  \end{aligned}
\end{equation}
Using this we can express conservation of energy as,
\begin{equation}
  \begin{aligned}
    \rho_0\frac{\partial E}{\partial t} &=
    \mathsf S^{ai}v_{a|i} - \mathsf S^{3i}v_{3|i} +
    \bm{\mathsf M}^{ij}\mathsf l_{3i|j} -
    v_{k|i}\tensor{\mathsf b}{^k_j}\bm{\mathsf M}^{ij} \\
    &= \mathsf S^{ij}v_{i|j} -
    \tensor{\mathsf b}{^i_k}\bm{\mathsf M}^{jk}v_{i|j} +
    \bm{\mathsf M}^{ij}\mathsf l_{3i|j} \\
    &= (\mathsf S^{ij} - \tensor{\mathsf b}{^i_k}\bm{\mathsf N}^{jk})v_{i|j} +
    \bm{\mathsf N}^{ij}\mathsf l_{3i|j}.
  \end{aligned}
\end{equation}
We recognise the term in the brackets as the modified Piola-Kirchhoff stress
tensor $\tilde{\mathsf S}$. Recalling that this is symmetric we can write
conservation of energy as,
\begin{equation}
  \rho_0\frac{\partial E}{\partial t} =
  \tilde{\mathsf S}^{ij}\mathsf n_{ij} +
  \bm{\mathsf N}^{ij}\mathsf l_{3i|j}.
\end{equation}
Using the kinematic results derived in \S\ref{sec:Kinematics} we can write this
as,
\begin{equation}
  \rho_0\frac{\partial E}{\partial t} =
  \tilde{\mathsf S}^{ij}\dot{\mathsf E}_{ij} +
  \bm{\mathsf N}^{ij}\dot{\mathsf H}_{ij},
\end{equation}
The fact that $\dot{\mathsf H}$ is symmetric means that only the symmetric part
of $\bm{\mathsf N}$ contributes to this expression, but as is discussed in
\S\ref{sec:BalanceLaws}, $\bm{\mathsf N}$ is assumed to be symmetric. Hence we
can write conservation of energy in component free form as,
\begin{equation}
  \rho_0\frac{\partial E}{\partial t} =
  \tr(\tilde{\mathsf S}\dot{\mathsf E}) +
    \tr(\bm{\mathsf N}\dot{\mathsf H}).
\end{equation}

%%%%%%%%%% Bibliography %%%%%%%%%%%%%%
\bibliographystyle{rspub_unsrt}
\bibliography{PhD}

\end{document}